# Linear tetramer formation in nonmagnetic pyrochlore niobate

Shota Nishida[1], Shunsuke Kitou[1,*], Shingo Toyoda[2], Yuiga Nakamura[3],
Yusuke Tokunaga[1], and Taka-hisa Arima[1,2]

[1]*Department of Advanced Materials Science, The University of Tokyo, Kashiwa 277-8561, Japan*
[2]*RIKEN Center for Emergent Matter Science (CEMS), Wako 351-0198, Japan*
[3]*Japan Synchrotron Radiation Research Institute (JASRI), SPring-8, Hyogo 679-5198, Japan*

**Abstract**
We investigate displacive short-range order in pyrochlore $Y_2Nb_2O_7$, which exhibits a nonmagnetic insulating state despite the presence of formally tetravalent $Nb^{4+}$ ($S = 1/2$) ions on the pyrochlore network. Synchrotron x-ray diffraction on a single crystal reveals a characteristic x-ray diffuse scattering (XDS) pattern primarily around $\boldsymbol{q} = \{0.5, 0.5, 2\}$. Reverse Monte Carlo (RMC) simulations uncover local Nb displacements along the $\langle 111 \rangle$ axes, leading to the formation of linear $Nb_4$ tetramers. Our findings highlight a crucial role of molecular orbital degrees of freedom in stabilizing the nonmagnetic insulating state. This study demonstrates that RMC analysis of XDS provides a powerful approach for elucidating short-range correlations and the underlying mechanisms governing the physical properties of crystalline materials.

**Introduction**

Many condensed-matter systems undergo phase transitions into lattice- and electron-ordered states at low temperatures, stabilized through interactions among various degrees of freedom. However, in certain materials, competing interactions prevent the formation of trivial order, leading instead to a *frustrated* state characterized by a complex energy landscape with multiple minima [1-3]. Such a multivalley energy structure allows external fields to induce switching between stable and metastable states, offering potential applications in memory devices, sensors, and information processing. For instance, the triangular-lattice magnet $CuFeO_2$ exhibits a rich phase diagram as a function of temperature and magnetic field [4], along with a nonlinear magnetic-field dependence of electric polarization. Moreover, a classical multivalley energy landscape can give rise to unconventional quantum ground states such as the resonating valence bond [5,6] and Kitaev quantum spin liquid states [7,8].

From this perspective, the pyrochlore niobate $R_2Nb_2O_7$ ($R$ = Y, Tb, Dy, Ho, Er, Tm, Yb, and Lu) represent an intriguing system [9-13], in which corner-sharing-tetrahedra of formally trivalent $R^{3+}$ and tetravalent $Nb^{4+}$ ions form an interpenetrating pyrochlore network [Fig. 1(a)]. Contrary to the simple expectation that each magnetic $Nb^{4+}$ ion donates one 4d valence electron ($S = 1/2$) [Figs. 1(b)-1(c)], Fukazawa and Maeno reported that $Y_2Nb_2O_7$, in which $Y^{3+}$ is a nonmagnetic ion, behaves as a nonmagnetic insulator over a wide temperature range between 0.35 and 800 K [10]. One may note here that several transition-metal compounds exhibit structural transitions accompanied by cluster formation—such as dimers in $VO_2$ [14,15], triangular trimers in $LiVO_2$ [16] and $GaNb_4Se_8$ [17], linear trimers in $Fe_3O_4$ [18], heptamer in $AlV_2O_4$ [19], and octamers in $CuIr_2S_4$ [20]. It is therefore presumable that Nb clusters such as dimers, tetramers, or hexamers may form in the pyrochlore niobate, giving rise to a spin-singlet ground state.

Previous in-house powder x-ray diffraction (XRD) data on $Y_2Nb_2O_7$ indicated a regular pyrochlore structure with the space group $Fd\bar{3}m$ [10], showing no evidence of Nb clustering. In contrast, powder neutron diffraction experiments on YCaNb$_2$O$_7$, in which half of the Y sites are replaced with Ca, revealed local Nb displacements along the threefold axes, possibly associated with molecular orbital formation induced by Nb–Nb bonding [13]. Theoretical calculations for Y$_2$Nb$_2$O$_7$ proposed a charge-singlet state accompanied by all-in/all-out type Nb displacements within Nb tetrahedra [21]. These findings suggest that charge and orbital degrees of freedom play a crucial role in frustrated pyrochlores, although a coherent microscopic picture has yet to be established.

In this study, we report synchrotron XRD experiments on single crystals of Y$_2$Nb$_2$O$_7$. Characteristic x-ray diffuse scattering (XDS) is observed primarily around $q$ = {0.5, 0.5, 2}. Reverse Monte Carlo (RMC) simulations identify that the XDS arises from correlated Nb displacements, associated with the nonmagnetic insulating nature of the system. These results uncover a short-range order driven by the molecular orbital degree of freedom in a geometrically frustrated pyrochlore lattice, pointing to an unexplored regime of electronic frustration.

**Methods**

A cylindrical rod-shaped single crystal of Y$_2$Nb$_2$O$_7$ was grown by a floating zone method from a thoroughly mixed stoichiometric ratio of Y$_2$O$_3$ and NbO$_2$ powders in argon gas. XRD measurements were performed on BL02B1 at a synchrotron facility SPring-8, Japan [22]. Crushed single crystals with dimensions of 60 × 50 × 40 μm$^3$ and 160 × 160 × 120 μm$^3$ were mounted on the tips of the glass rods and used for analyzing the average structure and disordered distortion, respectively. An N$_2$-gas-blowing device was employed to cool the crystals to 100 K. A two-dimensional CdTe PILATUS detector was used to record the XRD pattern. Incident x-ray energy was $E$ = 40 keV. The intensities of Bragg reflections of interplane distance $d$ > 0.28 Å were collected by CrysAlisPro program [23], in which the data were obtained by dividing the reciprocal space in increments of $\Delta\omega$ = 0.5°. Intensities of equivalent reflections were averaged and the structural parameters were refined to analyze the average structure by using Jana2006 [24]. The structural parameters were determined with high accuracy by performing a high-angle analysis utilizing the advantages of high-energy x-rays [25], where only reflections with sin$\theta/\lambda$ > 0.6 Å$^{-1}$ were used. RMC simulations to minimize the difference between the observed and calculated intensities of XDS were performed with the DISCUS program [26]. Details for the RMC analysis are explained in Supplemental Material (SM) [27].

**Results and discussion.**

XRD experiments on Y$_2$Nb$_2$O$_7$ reveal that the diffraction pattern does not significantly change between 100 and 300 K (Fig. S1 [27]). Structural refinement at 100 K assuming the space group $Fd\bar{3}m$ shows an anomalous elongation of the Nb atomic displacement parameters (ADPs) along the local threefold axes, as shown in Fig. 1(d). We further apply a site split model, in which each Nb site is split into two positions slightly displaced along the threefold axis with an occupancy of 1/2 [Fig. 1(e)]. This model significantly improves the agreement factor ($R$ value) from 15.16% to 5.89%, yielding a local Nb displacement of ±0.207(2) Å (see Tables S1-S3 for the details [27]).

Figure 2(a) shows the XRD data on the $h\ k\ 0$ plane at 100 K. Figures 2(b) and 2(c) present one-dimensional profiles of the XRD intensity along the $h$ and $k$ directions, respectively, corresponding to the lines labeled (i)-(iv) in Fig. 2(a). While sharp and intense Bragg peaks appear at positions where $h$ and $k$ are even integers satisfying $h + k = 4n$, broader and weaker XDS peaks are observed around half-integer positions. Figures 2(d) and 2(e) show the XRD data on the 6.5 $k\ l$ plane and a one-dimensional profile along the 6.5 4.5 $l$ line, respectively. Around 6.5 4.5 0, unlike in the $hk$ plane, side peaks appear on both sides of the central peak along the $l$ direction. Gaussian fitting reveals that the central and side XDS peaks correspond to the (6.5, 4.5, 0) and (6.5, 4.5, ±0.4) positions, respectively. The correlation length of the central XDS peak is estimated to be approximately $5.5a$ (≈ 56 Å) in all directions, whereas that of the side XDS peaks is approximately $4.6a$ (≈ 47 Å) along the $l$ direction (Fig. S3 [27]). The wave vector $q \approx$ {0.5, 0.5, 2} of the main XDS peak implies short-range correlations within one-dimensional Nb chains along the ⟨110⟩ axes, as shown in Fig. 2(f).

Based on the results of the structural analysis using the Nb site split model, the short-range order in $Y_2Nb_2O_7$ may arise primarily from Nb displacements. We hereafter consider only Nb displacements along the local threefold axes for simplicity. To investigate whether spin-singlet Nb dimers are formed, we perform Monte Carlo (MC) simulations using the Metropolis method [28] to model the Nb displacements based on the Pauling ice rule (2-in-2-out) [29], where only nearest neighbor (NN) Coulomb repulsive interactions are taken into account (see SM for the details [27]). The interaction energy is given by $E = \sum_{i,j:NN} 1/|r_j - r_i|$, where $r_i$ and $r_j$ represent the positions of the $i$-th and $j$-th Nb atoms on the pyrochlore lattice, respectively. The system size is set to $6a \times 6a \times 6a$ unit cells, corresponding to the correlation length estimated from the peak width of the XDS. Each Nb atom is displaced either parallel or antiparallel to the respective threefold axis by 0.2 Å. In the ground state, all Nb tetrahedra exhibit a 2-in-2-out displacement pattern that satisfies the ice rule (Fig. S4 [27]). In this displacive ice state, all Nb ions can form dimers giving rise to spin-singlet states, potentially accounting for the nonmagnetic nature of the system. However, the calculated XDS pattern based on this model [Fig. 3(b)] fails to reproduce the observed one shown in Fig. 3(a). Theoretically, if Nb atoms were displaced strictly following the ice rule, some pinch points should appear at {$4n+2$, 0, 0} [30-33], which is inconsistent with the experimental observation.

We also consider a tetrahedral-cluster model. If all Nb atoms were involved in the formation of tetrahedral molecules, Nb displacements of the so-called all-in/all-out type would emerge in the pyrochlore network. This distortion, corresponding to a breathing-type pyrochlore lattice, should develop without frustration, lowering the symmetry from $Fd\bar{3}m$ to $F\bar{4}3m$. However, no sharp reflections are observed on the $h\,k\,0$ plane with $h + k = 4n + 2$; instead, broad XDS peaks appear around half-integer positions. Furthermore, a simple calculation of the one-electron energy levels for a regular Nb tetramer predicts a lowest-lying singlet ($a_1$) orbital, followed by a doublet (e) and a triplet ($t_2$), similar to the case of $GaNb_4Se_8$ [17]. Four 4d electrons in the Nb tetramer would therefore behave as $S = 1$ rather than $S = 0$.

To elucidate the short-range order of Nb displacements along the local threefold axes in $Y_2Nb_2O_7$, we perform RMC simulations in $6a \times 6a \times 6a$ unit cells using two models. The first is a site-split Ising-type model, in which each Nb atom is displaced by ±0.207 Å from the ideal pyrochlore positions. The second, hereafter referred to as the "distribution model," assumes Nb displacement distributions consistent with the average structural analysis shown in Fig. 3(c). Figure 3(d) presents the XDS pattern on the $h\,k\,0$ plane obtained from the RMC simulation based on the latter model, which successfully reproduces the observed pattern (see also Fig. S9 [27]). It is worth noting that the RMC analysis result based on the Ising model shows poorer agreement with the experimental data than that obtained from the distribution model (Fig. S8 [27]). Short-range order arising from atomic displacements of transition-metal ions has also been observed in another pyrochlore oxide $Y_2Mo_2O_7$ [34,35], where local Mo–Mo dimer formation is associated with the spin-glass state.

Figure 4(a) presents the displacement pattern obtained from the RMC analysis in the $6a \times 6a \times 6a$ supercell. The colors of the $Nb_4$ tetrahedra represent five distinct distortion modes. If the displacive ice rule were satisfied in $Y_2Nb_2O_7$, all $Nb_4$ tetrahedra would exhibit the blue-colored 2-in-2-out distortion mode. However, the RMC analysis suggests that five tetrahedral distortion modes are present with their probabilities close to those expected for random Ising-type displacements (Table I). We further examine two-dimensional Nb planes that contribute significantly to the XDS intensity (Fig. S10 [27]). Figure 4(b) shows Nb displacements in an *ab*-plane, where one-dimensional Nb$_4$ tetramers are formed along the [110] axis. Similar linear tetramers are observed in the structure obtained from the RMC analysis. A total of 226 tetramers oriented perpendicular to the *c*-axis are identified, involving approximately one-fourth of Nb atoms.

It is useful to discuss a hypothetical long-range ordered structure characterized by linear tetramers running along the ⟨110⟩ axes (Fig. S12 [27]). The unit cell becomes $\sqrt{2}a \times \sqrt{2}a \times a$ with the tetragonal space group $P\bar{4}m2$. This structure roughly reproduces the experimental XDS pattern, although it generates an excessive number of superlattice peaks. The numbers of tetrahedral distortion modes are consistent with those expected for a completely random Ising case (Table I). However, the actual crystal does not form the long-range ordered structure, which may be related to the presence of distinct short-range order corresponding to $\boldsymbol{q} \approx \{0.5, 0.5, 2\pm0.4\}$, which cannot be captured within the present RMC analysis.

Finally, we discuss the relationship between the observed short-range ordered structure and the nonmagnetic insulating nature of Y$_2$Nb$_2$O$_7$. Each Nb$^{4+}$ ion at the trigonally compressed octahedral site hosts one electron in the a$_{1g}$ orbital [Fig. 1(c)]. The a$_{1g}$ orbitals on four Nb sites within a linear tetramer reorganize into four molecular orbitals with different energies, as shown in Fig. 4(c), analogous to those in a butane molecule. The lowest and second-lowest σ$_1$ and σ$_2$ orbitals each accommodate two electrons. If the σ$_2$ and σ$_3$ bands are separated in energy, the system can behave as a nonmagnetic band insulator. The linear tetramers can align along six equivalent ⟨110⟩ axes, which likely hinders the development of long-range order. This structural degeneracy may be responsible for the formation of the observed short-range order. Such linear metal clusters on pyrochlore lattice are unprecedented. External perturbations, such as pressures or magnetic fields, may tune this delicate balance between electronic and structural states, potentially leading to novel electronic phases driven by hidden degrees of freedom.

In summary, we investigated the short-range ordered structure of the nonmagnetic insulator Y$_2$Nb$_2$O$_7$. Synchrotron XRD combined with RMC simulations revealed that Nb displacements along the local threefold axes form linear Nb$_4$ tetramers, giving rise to short-range order. This behavior is closely related to the nonmagnetic insulating nature of the system and suggests that the resultant molecular-orbital reconstruction plays a crucial role in stabilizing this state.

*Acknowledgements.* This work was supported by JSPS KAKENHI (Grant Nos. 24K17006 and 24H01644), and JST FOREST (Grant No. JPMJFR2362). The synchrotron radiation experiments were performed at SPring-8 with the approval of the Japan Synchrotron Radiation Research Institute (JASRI) (Proposal 2023B1603 and 2025A1505).

**References**

1. R. Moessner, and A. P. Ramirez. Geometrical frustration. Physics Today **59**, 24-29 (2006).
2. D. A. Keen, and A. L. Goodwin, The crystallography of correlated disorder. Nature **521**, 303-309 (2015).
3. A. Simonov, and A. L. Goodwin, Nat. Rev. Chem. **4**, 657-673 (2020).
4. T. Kimura, J. C. Lashley, and A. P. Ramirez, Inversion-symmetry breaking in the noncollinear magnetic phase of the triangular-lattice antiferromagnet CuFeO$_2$. Phys. Rev. B **73**, 220401(R) (2006).
5. L. C. Pauling, A resonating-valence-bond theory of metals and intermetallic compounds. Proc. R. Soc. London, Ser. A **196**, 343 (1949).

6. P. W. Anderson, Resonating valence bonds: A new kind of insulator" Mater. Res. Bull. **8**, 153-160 (1973).
7. A. Kitaev, Anyons in an exactly solved model and beyond. Ann. Phys. **321**, 2-111 (2006).
8. H. Takagi, T. Takayama, G. Jackeli, G. Khaliullin, and S. E. Nagler, Concept and realization of Kitaev quantum spin liquids. Nat. Rev. Phys. **1**, 264-280 (2019).
9. M. A. Subramanian, G. Aravamudan and G. V. Subba Rao, Oxide pyrochlores—A review. Prog. Solid State Chem. **15**, 55-143 (1983).
10. H. Fukazawa, and Y. Maeno, Magnetic ground state of the pyrochlore oxide $Y_2Nb_2O_7$. Phys. Rev. B **67**, 054410 (2003).
11. Y. M. Jana, O. Sakai, R. Higashinaka, H. Fukazawa, Y. Maeno, P. Dasgupta, and D. Ghosh, Spin-glass-like magnetic ground state of the geometrically frustrated pyrochlore niobate $Tb_2Nb_2O_7$. Phys. Rev. B **68**, 174413 (2003).
12. O. Sakai, Y. Jana, R. Higashinaka, H. Fukazawa, S. Nakatsuji, and Y. Maeno, New Compounds Based on Pyrochlore Structure: $R_2Nb_2O_7$ ($R$ = Dy, Yb). J. Phys. Soc. Jpn. **73**, 2829-2833 (2004).
13. T. M. McQueen, D. V. West, B. Muegge, Q. Huang, K. W. Noble, H. W. Zandbergen, and R. J. Cava, Frustrated ferroelectricity in niobate pyrochlores. J. Phys.: Condens. Matter **20**, 235210 (2008).
14. G. Andersson, Studies on Vanadium Oxides. Acta Chim. Scand. **10**, 623 (1956).
15. S. Kitou, A Nakano, M Imaizumi, Y Nakamura, I Terasaki, and T. Arima, Molecular orbital formation and metastable short-range ordered structure in $VO_2$. Phys. Rev. B **109**, L100101 (2024).
16. J. B. Goodenough, G. Dutta, and A. Manthiram, Lattice instabilities near the critical V-V separation for localized versus itinerant electrons in $LiV_{1-y}M_yO_2$ (M = Cr or Ti) $Li_{1-x}VO_2$. Phys. Rev. B **43**, 10170 (1991).
17. S. Kitou, M. Gen, Y. Nakamura, Y. Tokunaga, and T. Arima, Cluster Rearrangement by Chiral Charge Order in Lacunar Spinel $GaNb_4Se_8$. Chem. Mater. **36**, 2993 (2024).
18. M. S. Senn, J. P. Wright, and J. P. Attfield, Charge order and three-site distortions in the Verwey structure of magnetite. Nature **481**, 173 (2012).
19. Y. Horibe, M. Shingu, K. Kurushima, H. Ishibashi, N. Ikeda, K. Kato, Y. Motome, N. Furukawa, S. Mori, and T. Katsufuji, Spontaneous Formation of Vanadium "Molecules" in a Geometrically Frustrated Crystal: $AlV_2O_4$. Phys. Rev. Lett. **96**, 086406 (2006).
20. P. G. Radaelli, Y. Horibe, M. J. Gutmann, H. Ishibashi, C. H. Chen, R. M. Ibberson, Y. Koyama, Y.-S. Hor, V. Kiryukhin, and S.-W. Cheong, Formation of isomorphic $Ir^{3+}$ and $Ir^{4+}$ octamers and spin dimerization in the spinel $CuIr_2S_4$. Nature **416**, 155-158 (2002).
21. P. Blaha, D. J. Singh, and K. Schwarz, Geometric Frustration, Electronic Instabilities, and Charge Singlets in $Y_2Nb_2O_7$. Phys. Rev. Lett. **93**, 216403 (2004).
22. K. Sugimoto, H. Ohsumi, S. Aoyagi, E. Nishibori, C. Moriyoshi, Y. Kuroiwa, H. Sawa, M. Takata, Extremely High Resolution Single Crystal Diffractometry for Orbital Resolution using High Energy Synchrotron Radiation at SPring-8. AIP Conf. Proc. **1234**, 887 (2010).
23. CrysAlisPro, Agilent Technologies Ltd, Yarnton, Oxfordshire, England, (2014).
24. V. Petříček, M. Dušek, L. Palatinus, Crystallographic Computing System JANA2006: General features. Z. Kristallogr. Cryst. Mater. **229**, 345 (2014).
25. S. Kitou, T. Fujii, T. Kawamoto, N. Katayama, S. Maki, E. Nishibori, K. Sugimoto, M. Takata, T. Nakamura, H. Sawa, Successive Dimensional Transition in $(TMTTF)_2PF_6$ Revealed by Synchrotron X-ray Diffraction. Phys. Rev. Lett. **119**, 065701 (2017).

26. T. Proffen and R. B. Neder, DISCUS: a program for diffuse scattering and defect-structure simulation. J. Appl. Cryst. **30**, 171 (1997).
27. See Supplemental Material for details of the structural and XDS analysis.
28. N. Metropolis, A. W. Rosenbluth, M. N. Rosenbluth, and A. H. Teller, J. Chem. Phys. **21**, 1087 (1953).
29. L. Pauling, The structure and entropy of ice and of other crystals with some randomness of atomic arrangement. J. Am. Chem. Soc. **57** 2680-2684 (1935).
30. S. T. Bramwell, M. J. Harris, B. C. den Hertog, M. J. P. Gingras, J. S. Gardner, D. F. McMorrow, A. R. Wildes, A. L. Cornelius, J. D. M. Champion, R. G. Melko, and T. Fennell, Spin Correlations in $Ho_2Ti_2O_7$: A Dipolar Spin Ice System. Phys. Rev. Lett. **87**, 047205 (2001).
31. T. Fennell, O. A. Petrenko, B. Fåk, S. T. Bramwell, M. Enjalran, T. Yavors'kii, M. J. P. Gingras, and G. Balakrishnan, Neutron scattering investigation of the spin ice state in $Dy_2Ti_2O_7$. Phys. Rev. B 70, 134408 (2004).
32. D. J. P. Morris, D. A. Tennant, S. A. Grigera, B. Klemke, C. Castelnovo, R. Moessner, C. Czternasty, M. Meissner, K. C. Rule, J.-U. Hoffmann, K. Kiefer, S. Gerischer, D. Slobinsky, and R. S. Perry, Dirac Strings and Magnetic Monopoles in the Spin Ice $Dy_2Ti_2O_7$. Science **326**, 411 (2009).
33. T. Fennell, P. P. Deen, A. R. Wildes, K. Schmalzl, D. Prabhakaran, A. T. Boothroyd, R. J. Aldus, D. F. McMorrow, and S. T. Bramwell, Magnetic Coulomb Phase in the Spin Ice $Ho_2Ti_2O_7$. Science **326**, 415 (2009).
34. J. E. Greedan, D. Gout, A. D. Lozano-Gorrin, S. Derakhshan, Th. Proffen, H.-J. Kim, E. S. Bozin, and S. J. L. Billinge, Local and average structures of the spin-glass pyrochlore $Y_2Mo_2O_7$ from neutron diffraction and neutron pair distribution function analysis. Phys. Rev. B **79**, 014427 (2009).
35. P. M. M. Thygesen, J. A. M. Paddison, R. Zhang, K. A. Beyer, K. W. Chapman, H. Y. Playford, M. G. Tucker, D. A. Keen, M. A. Hayward, and A. L. Goodwin, Orbital Dimer Model for the Spin-Glass State in $Y_2Mo_2O_7$. Phys. Rev. Lett. **118**, 067201 (2017).

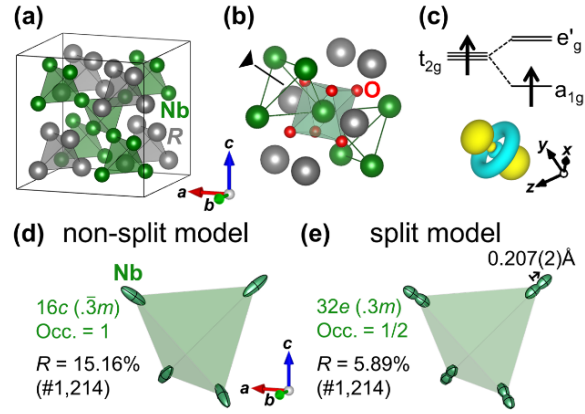

FIG. 1. (a) Crystal structure of $R_2Nb_2O_7$. The $R$ and Nb atoms each form pyrochlore networks. (b) An Nb atom is coordinated by six O atoms, forming an $NbO_6$ octahedron. This octahedron is contracted along the three-fold rotation axis, indicated by a solid triangle. (c) Schematic of the $t_{2g}$ orbital state of a $Nb^{4+}$ ion with one 4d electron and the wavefunction of the $a_{1g}$ orbital. Structural analysis of $Y_2Nb_2O_7$ at 100 K based on 1,214 equivalent reflections for (d) the non-split and (e) split models of the Nb sites, assuming the $Fd\bar{3}m$ space group. Nb atomic displacement parameters are shown as ellipsoids with a 50% probability.

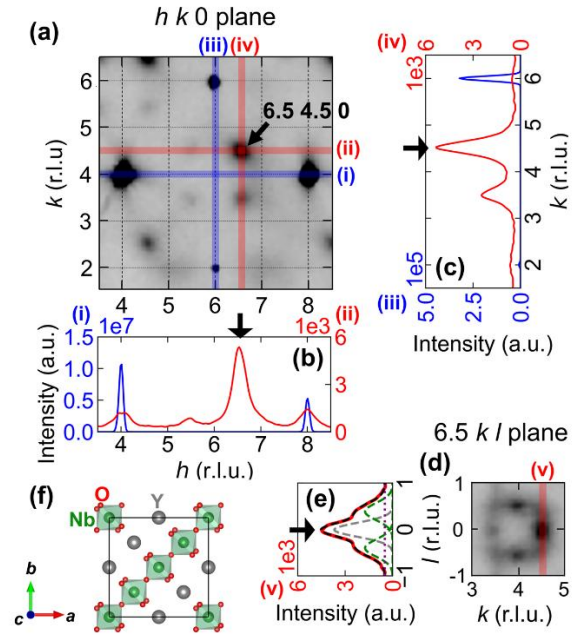

FIG. 2. (a) XRD data on the $h\,k\,0$ plane of $Y_2Nb_2O_7$ at 100 K. (b) Blue and red lines indicate the intensity profiles along the $h\,4\,0$ and $h\,4.5\,0$ lines, respectively. (c) Blue and red lines indicate the intensity profiles along the $6\,k\,0$ and $6.5\,k\,0$ lines, respectively. (d) XRD data on the $6.5\,k\,l$. (e) A red line indicates the intensity profile along the $6.5\,4.5\,l$ line. A black dashed line represents the fitting result using three Gaussian functions: one gray dashed line for the central peak and two green dashed lines for the side peaks. (f) Crystal structure of a single layer in the $ab$ plane of $Y_2Nb_2O_7$. $NbO_6$ octahedra form one-dimensional chains along the $\langle 110 \rangle$ directions.

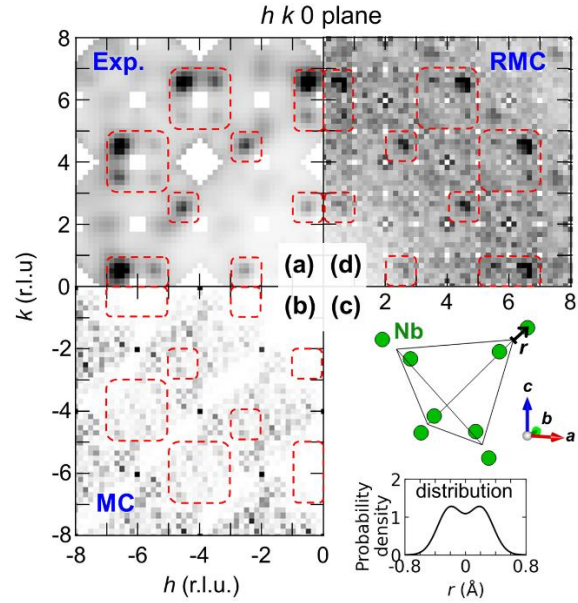

FIG. 3. (a) XRD data on the $h\,k\,0$ plane of $Y_2Nb_2O_7$ at 100 K, with Bragg peak intensities removed. (b) Diffuse scattering pattern on the $h\,k\,0$ plane calculated by the MC simulation based on the displacive ice state. (c) Schematic of Nb displacements used in the RMC simulation based on the distribution model. The displacement directions are constrained along the $\langle 111 \rangle$ axes, and the displacement magnitudes are determined based on the ADPs obtained by the structural analysis. (d) Diffuse scattering pattern on the $h\,k\,0$ plane calculated by the RMC simulation based on the distribution model in (c).

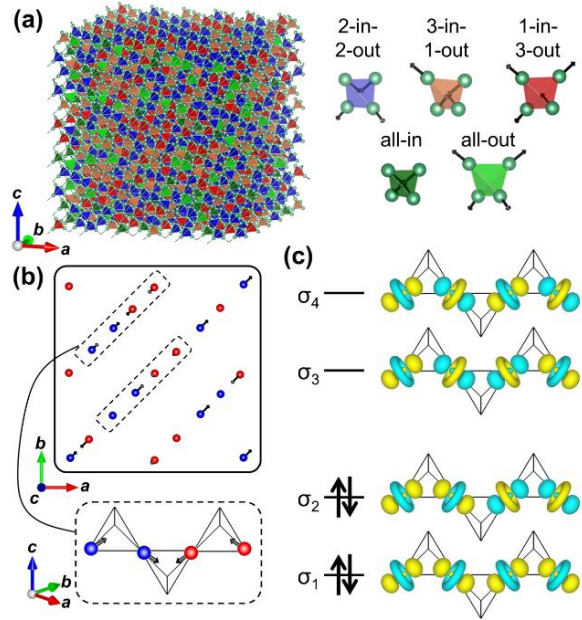

FIG. 4. (a) Crystal structure showing only Nb atoms within a 6$a$ × 6$a$ × 6$a$ supercell, obtained from RMC simulations. Blue, orange, red, dark green, and light green tetrahedra represent five distinct distortion modes. (b) Nb displacements in a part of the single layer in the $ab$ plane. The blue and red colors of the atoms indicate the displacement directions, either [+1+1±1] or [-1-1±1]. The Nb displacements are displayed as black vectors, scaled by a factor of five. (c) Schematic of molecular orbitals of the one-dimensional Nb$_4$ tetramer. Yellow and light blue represent the positive and negative signs of the wavefunctions, respectively. The four molecular orbitals are labeled as $\sigma_1$, $\sigma_2$, $\sigma_3$, and $\sigma_4$ in order of increasing energy. The $\sigma_1$ and $\sigma_2$ orbitals each contain two electrons, which are paired in a singlet state.

Table I. The number of each distortion mode—MC (displacive ice), RMC (distribution), and Random Ising—for the Nb$_4$ tetrahedron. The numbers in parentheses for the Random Ising model indicate the relative ratio of each mode.

| Analysis model | 2-in-2-out | 3-in-1-out | 1-in-3-out | all-in | all-out |
|---|---|---|---|---|---|
| MC | 1728 | 0 | 0 | 0 | 0 |
| RMC | 638 | 409 | 414 | 134 | 131 |
| Random Ising | 648 (6) | 432 (4) | 432 (4) | 108 (1) | 108 (1) |

# Supplemental Material for
# Linear tetramer formation in nonmagnetic pyrochlore niobate

Shota Nishida, Shunsuke Kitou, Shingo Toyoda, Yuiga Nakamura,
Yusuke Tokunaga, and Taka-hisa Arima

XRD experiments

Figure S1 shows the x-ray diffraction (XRD) data of $Y_2Nb_2O_7$ at 300 and 100 K, revealing no clear temperature dependence in the x-ray diffuse scattering (XDS). Figure S2 shows the temperature dependence of the crystal structure of $Y_2Nb_2O_7$, assuming the space group $Fd\bar{3}m$, which shows no significant temperature dependence. The structural analysis results are summarized in Tables S1-S3. Figure S3 shows the Gaussian fitting results of the XDS around 6.5 4.5 0. The full width at half maximum (FWHM) of the XDS peak around 6.5 4.5 0 is obtained as 0.376(7), 0.385(5), and 0.355(11) along the $h$, $k$, and $l$ directions, respectively. The correlation length $\xi$ is estimated as $\xi = 2a/\text{FWHM}$.

XDS intensity

The XDS intensity can be evaluated from deviations of the short-range ordered structure from the periodic long-range ordered structure. The structural factor of a crystal can be expressed as

$$F(\mathbf{K}) = \langle F(\mathbf{K}) \rangle + [F(\mathbf{K}) - \langle F(\mathbf{K}) \rangle]$$
$$= \langle F(\mathbf{K}) \rangle + F_D(\mathbf{K}), \qquad (1)$$

where $\langle F(\mathbf{K}) \rangle$ represents the average structure factor corresponding to the long-range periodic order at the scattering vector $\mathbf{K}$, $F(\mathbf{K})$ denotes the total structure factor that includes local atomic displacements, and $F_D(\mathbf{K})$ is the diffuse component of the structure factor. The XDS intensity $I_D(\mathbf{K})$ is then given by the square magnitude of the diffuse structure factor:

$$I_D(\mathbf{K}) \propto |F_D(\mathbf{K})|^2$$
$$= |F(\mathbf{K}) - \langle F(\mathbf{K}) \rangle|^2. \qquad (2)$$

MC simulation

To realize a displacive ice state in $Y_2Nb_2O_7$, we performed Monte Carlo (MC) simulations using the Metropolis method [1] to model the Nb displacements, considering only nearest-neighbor Coulomb repulsive interactions. As a result, the Nb atoms adopt the 2-in-2-out displacement pattern in all tetrahedra [Fig. S4(a)], thereby realizing the displacive ice state, with six distinct types of 2-in-2-out configurations [Fig. S4(b)]. Figure S4(c) shows the calculated XDS pattern on the $h\,k\,0$ plane.

RMC analysis

Reverse Monte Carlo (RMC) simulations to minimize the difference between the observed and calculated intensities of XDS were performed with the DISCUS program [2]. For the XDS analysis, the background intensity from air scattering was subtracted using the diffraction data of the glass rods measured under the same conditions. The XRD data in the range of $-8 \leq h, k, l \leq 8$ were used for the analysis, excluding the Bragg peak intensities. The observed and calculated XRD intensities were averaged according to the XDS symmetry of the point group $m\bar{3}m$ (Fig. S5). To set the voxel size in the three-dimensional reciprocal space to $(1/6\text{ r.l.u.})^3$, the two-dimensional plane images for the analysis covered a thickness of 1/6 r.l.u. For example, the $h\,k\,1$ plane was prepared as an average of the $h\,k\,0.917$ to $h\,k\,1.083$ reciprocal planes. Considering the correlation length estimated from the peak width of the XDS [Fig. S3], the calculation system size was set to $6a \times 6a \times 6a$ unit cells. Figures S6 and S7 show the schematic of the RMC simulations and the RMC cycle dependence of the $\sigma$, $\chi^2$, and $f$, respectively. Here, $\sigma$ is a modeling parameter that controls the fraction of bad moves accepted, $\chi^2$ is a goodness-of-fit parameter, and $f$ is a scale factor, as explained in [2].

Figure S8 shows the RMC analysis results, in which two distinctive approaches—based on distribution and Ising models—are applied. In both models, the Nb displacement directions are constrained to be parallel or antiparallel to the local threefold axes. In the Ising model, the displacement magnitude is fixed at 0.207 Å, consistent with the Nb site-split model analysis result. In contrast, the distribution model employs two Gaussian distributions derived from the atomic displacement parameters of Nb atoms were used [Fig. 3(c)]. The analysis based on the Ising model reproduced the positions of the XDS peaks to some extent, but failed to accurately capture the intensity distribution [Figs. S8(e) and S8(f)]. In contrast, the distribution model successfully reproduced both the peak positions and the overall intensity distribution of the XDS [Figs. S8(c) and S8(d)]. The superior agreement of the distribution model with the experimental data, compared to the Ising model, is attributed to the presence of Nb$_4$ tetrahedra exhibiting a variety of displacement modes in the RMC analysis results. Figure S9 shows experimental and simulated XDS patterns on the $hk$ planes for different $l$ values, exhibiting good agreement across all planes.

From the results of the RMC simulations based on the distribution model, we identified two-dimensional planes within the $6a \times 6a \times 6a$ supercell that closely reproduce the experimental XDS pattern, as shown in Fig. S10(c). Figure S10(a) displays the $6a \times 6a \times a$ structure viewed along the $c$-axis, where several linear Nb$_4$ tetramers are observed [Fig. S10(b)].

XRD intensity calculation based on the long-range ordered model

We investigated the long-range ordered structure based on the Ising-type Nb displacements that form Nb$_4$ tetramers along the $\langle 110 \rangle$ directions, as shown in Fig. S11. As a result, we found that a structure with a $\pi$ phase shift between the one-dimensional chains **(4)** reproduced the experimental data relatively well. Such two-dimensional Nb displacements can form a long-range ordered structure in a $\sqrt{2}a \times \sqrt{2}a \times a$ supercell with the tetragonal space group $P\bar{4}m2$, as shown in Fig. S12. This structure roughly reproduces the experimental XRD pattern, though with an excessive number of superlattice peaks.

Table S1. Summary of crystallographic data of $Y_2Nb_2O_7$ at 100 K.

| | |
|---|---|
| Wavelength (Å) | 0.30925 |
| Crystal dimension (μm$^3$) | 60 × 50 × 40 |
| Space group | $Fd\bar{3}m$ |
| $a$ (Å) | 10.2652(7) |
| Z | 8 |
| $F(000)$ | 1728 |
| $(\sin\theta/\lambda)_{max}$ (Å$^{-1}$) | 1.79 |
| Number of total reflections | 27790 |
| Average redundancy | 21.728 |
| Completeness (%) | 98.16 |
| Number of unique reflections ($\sin\theta/\lambda > 0.6$ Å$^{-1}$ / all) | 999 / 1214 |
| Non Nb site split model | |
| $N_{parameters}$ | 10 |
| $R_1$ ($\sin\theta/\lambda > 0.6$ Å$^{-1}$ / all) (%) | 15.16 / 15.70 |
| $wR_2$ ($\sin\theta/\lambda > 0.6$ Å$^{-1}$ / all) (%) | 17.60 / 17.65 |
| GOF ($\sin\theta/\lambda > 0.6$ Å$^{-1}$ / all) | 11.36 / 10.32 |
| Nb site split model | |
| $N_{parameters}$ | 11 |
| $R_1$ ($\sin\theta/\lambda > 0.6$ Å$^{-1}$ / all) (%) | 5.89 / 6.62 |
| $wR_2$ ($\sin\theta/\lambda > 0.6$ Å$^{-1}$ / all) (%) | 11.18 / 11.25 |
| GOF ($\sin\theta/\lambda > 0.6$ Å$^{-1}$ / all) | 7.22 / 6.58 |

Table S2. Structural parameters of Y$_2$Nb$_2$O$_7$ at 100 K, assuming the non Nb site split model.

| Atom | Wyckoff position | Site symmetry | $x$ | $y$ | $z$ | Occupancy |
|---|---|---|---|---|---|---|
| Y | 16$d$ | .$\bar{3}m$ | 1/2 | 1/2 | 1/2 | 1 |
| Nb | 16$c$ | .$\bar{3}m$ | 0 | 0 | 0 | 1 |
| O(1) | 48$f$ | 2.$mm$ | 0.3372(5) | 1/8 | 1/8 | 1 |
| O(2) | 8$b$ | $\bar{4}3m$ | 3/8 | 3/8 | 3/8 | 1 |

| Atom | $U_{11}$ (10$^{-2}$ Å$^2$) | $U_{22}$ (10$^{-2}$ Å$^2$) | $U_{33}$ (10$^{-2}$ Å$^2$) | $U_{12}$ (10$^{-2}$ Å$^2$) | $U_{13}$ (10$^{-2}$ Å$^2$) | $U_{23}$ (10$^{-2}$ Å$^2$) |
|---|---|---|---|---|---|---|
| Y | 0.868(9) | = $U_{11}$ | = $U_{11}$ | -0.180(3) | = $U_{12}$ | = $U_{12}$ |
| Nb | 3.20(3) | = $U_{11}$ | = $U_{11}$ | 2.26(4) | = $U_{12}$ | = $U_{12}$ |
| O(1) | 3.8(2) | 1.34(4) | = $U_{22}$ | 0 | 0 | 0.22(5) |
| O(2) | 0.88(4) | = $U_{11}$ | = $U_{11}$ | 0 | 0 | 0 |

Table S3. Structural parameters of Y$_2$Nb$_2$O$_7$ at 100 K, assuming the Nb site split model.

| Atom | Wyckoff position | Site symmetry | $x$ | $y$ | $z$ | Occupancy |
|---|---|---|---|---|---|---|
| Y | 16$d$ | .$\bar{3}m$ | 1/2 | 1/2 | 1/2 | 1 |
| Nb | 32$e$ | .3$m$ | 0.01165(7) | = $x$ | = $x$ | 1/2 |
| O(1) | 48$f$ | 2.$mm$ | 0.3370(2) | 1/8 | 1/8 | 1 |
| O(2) | 8$b$ | $\bar{4}3m$ | 3/8 | 3/8 | 3/8 | 1 |

| Atom | $U_{11}$ (10$^{-2}$ Å$^2$) | $U_{22}$ (10$^{-2}$ Å$^2$) | $U_{33}$ (10$^{-2}$ Å$^2$) | $U_{12}$ (10$^{-2}$ Å$^2$) | $U_{13}$ (10$^{-2}$ Å$^2$) | $U_{23}$ (10$^{-2}$ Å$^2$) |
|---|---|---|---|---|---|---|
| Y | 0.822(6) | = $U_{11}$ | = $U_{11}$ | -0.2026(19) | = $U_{12}$ | = $U_{12}$ |
| Nb | 1.507(12) | = $U_{11}$ | = $U_{11}$ | 0.570(11) | = $U_{12}$ | = $U_{12}$ |
| O(1) | 2.57(9) | 1.24(3) | = $U_{22}$ | 0 | 0 | 0.312(3) |
| O(2) | 0.81(2) | = $U_{11}$ | = $U_{11}$ | 0 | 0 | 0 |

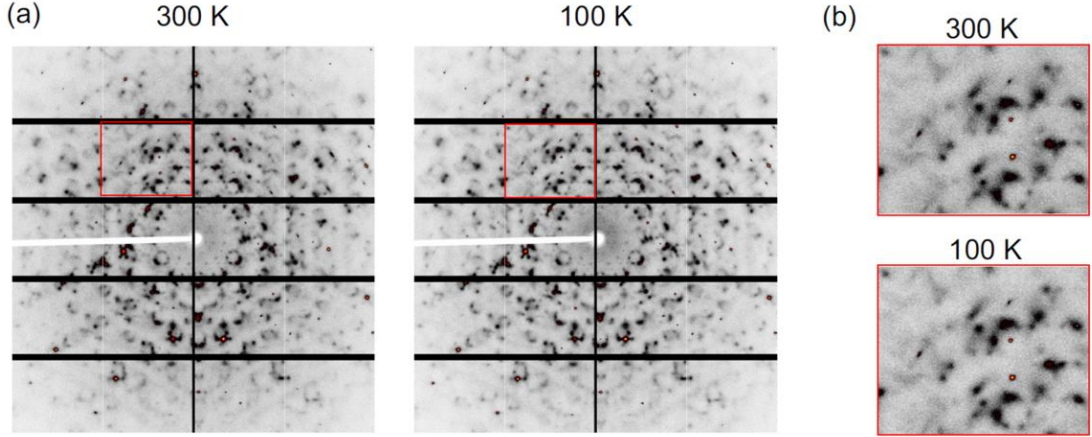

FIG. S1. (a) XRD data of $Y_2Nb_2O_7$ at 300 and 100 K. (b) Enlarged views of the red squares in (a).

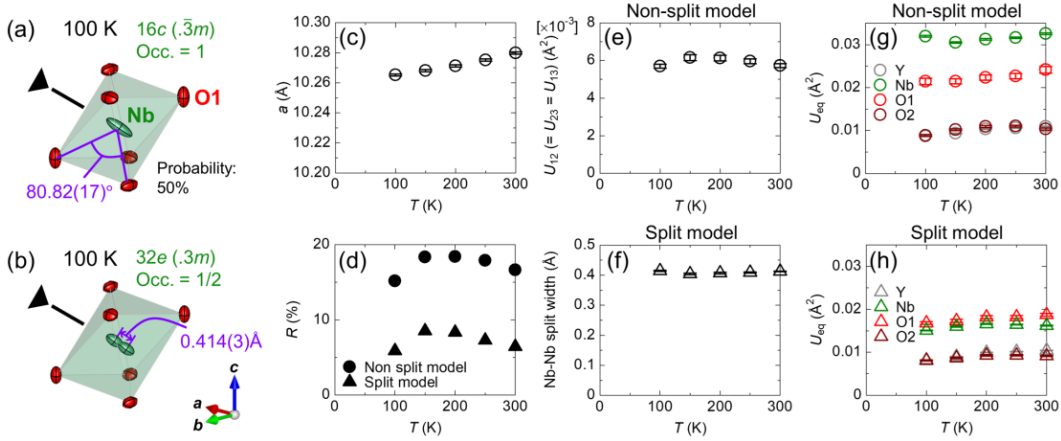

FIG. S2. Structural analysis results around a $NbO_6$ octahedron for the (a) non-split and (b) split models of the Nb sites, assuming the $Fd\bar{3}m$ space group. Atomic displacement parameters are shown as ellipsoids with a 50% probability. (c)-(h) Temperature dependence of the lattice constant $a$, $R$ value, atomic displacement parameter $U_{12}$ of Nb (non-split model), Nb–Nb split width (split model), and $U_{eq}$ [$= (U_{11} + U_{22} + U_{33})/3$] (non-split and split models) of $Y_2Nb_2O_7$.

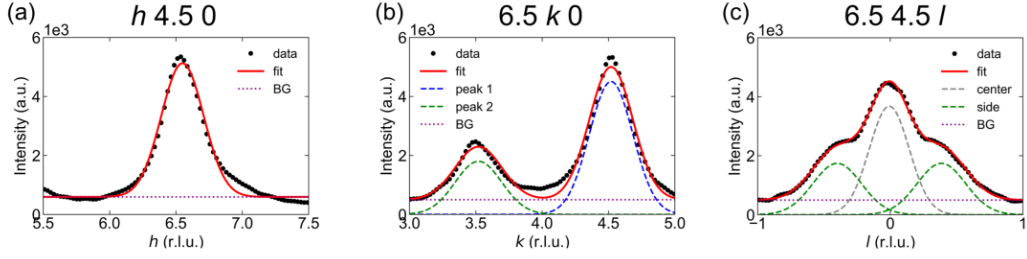

FIG. S3. Gaussian fitting for the XDS intensity around 6.5 4.5 0 at 100 K, where (a), (b), and (c) represent the $h$ 4.5 0, 6.5 $k$ 0, and 6.5 4.5 $l$ lines, respectively.

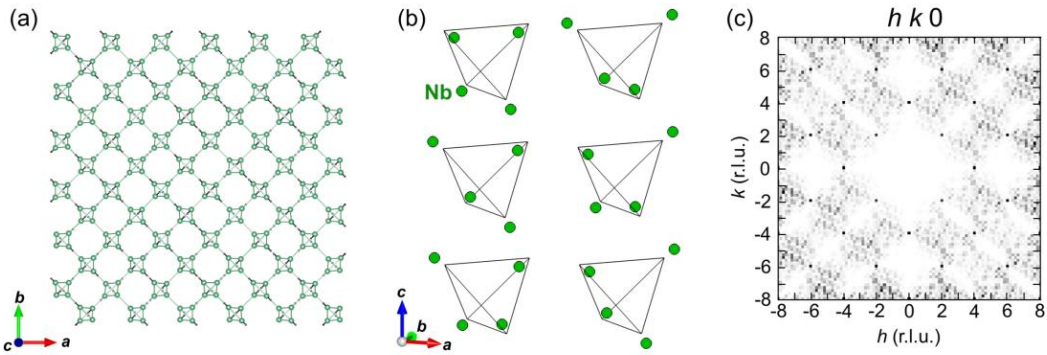

FIG. S4. (a) Results of the MC simulations in a $6a \times 6a \times 6a$ supercell. The Nb displacements, shown by black arrows, realize the displacive ice state. (b) Six distinct types of 2-in-2-out configurations on energetically degenerate $Nb_4$ tetrahedra. (c) Calculated XDS pattern on the $h\,k\,0$ plane.

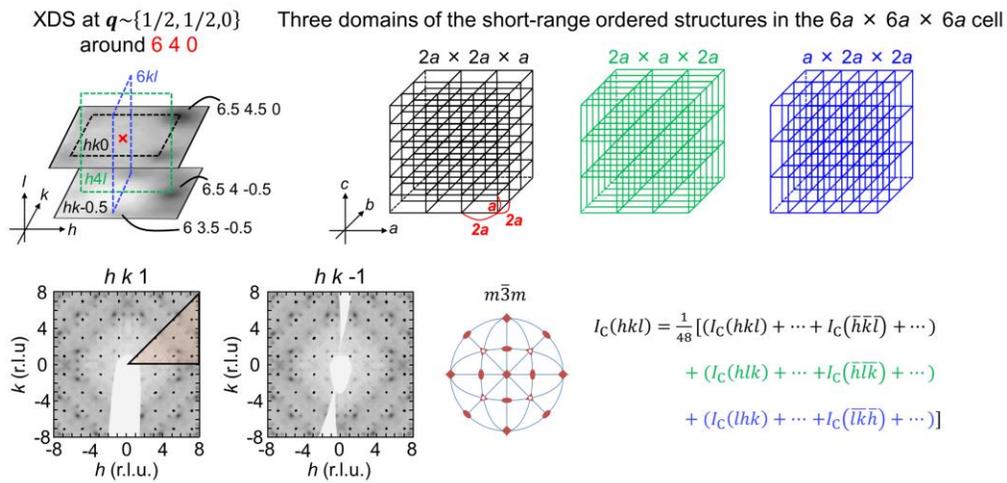

FIG. S5. Domain structures of the short-range order and symmetrized XRD data for the RMC simulations.

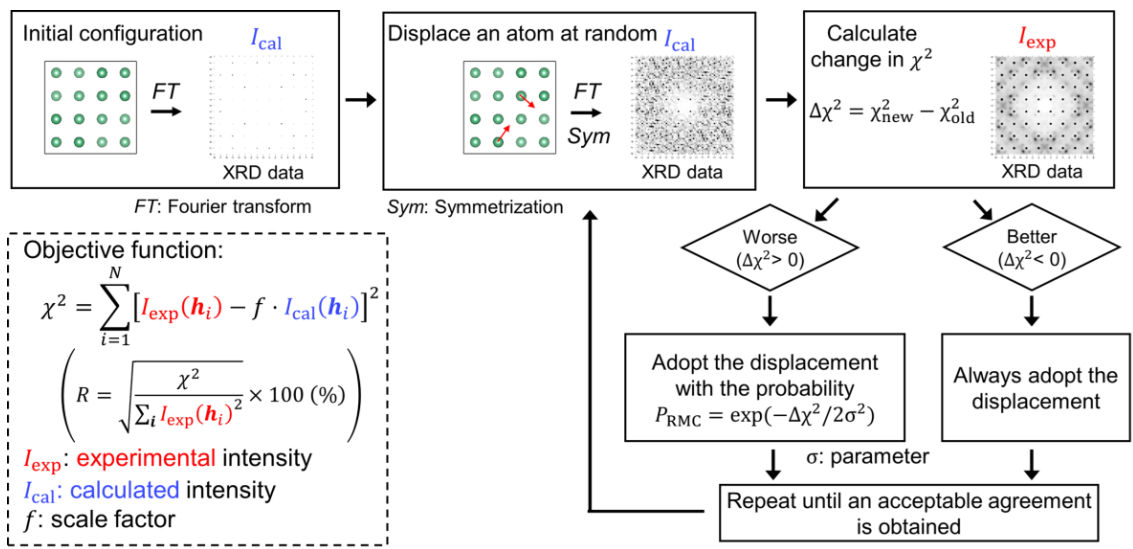

FIG. S6. Schematic of the RMC simulations.

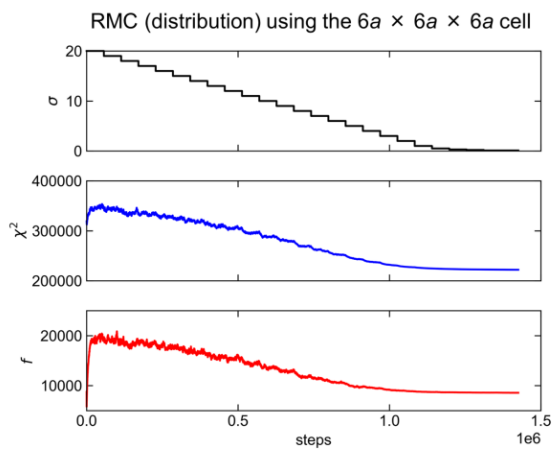

FIG. S7. RMC cycles using the distribution model. $\sigma$ is a modeling parameter that controls the fraction of bad moves accepted. $\chi^2$ is a goodness-of-fit parameter, and $f$ is a scale factor.

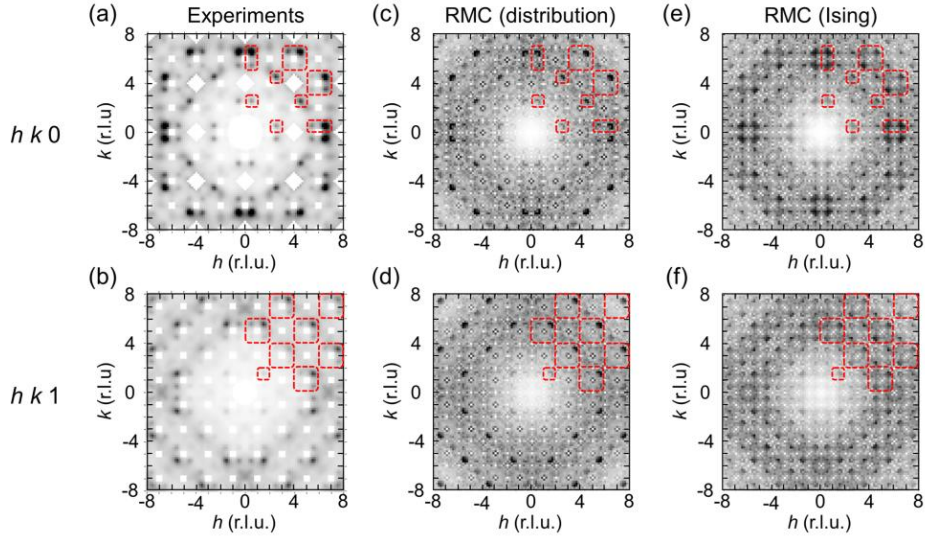

FIG. S8. (a),(b) XRD data on the $h\,k\,0$ and $h\,k\,1$ planes of $Y_2Nb_2O_7$ at 100 K, with Bragg peak intensities removed. Results of the RMC simulations using (c),(d) distribution and (e),(f) Ising models.

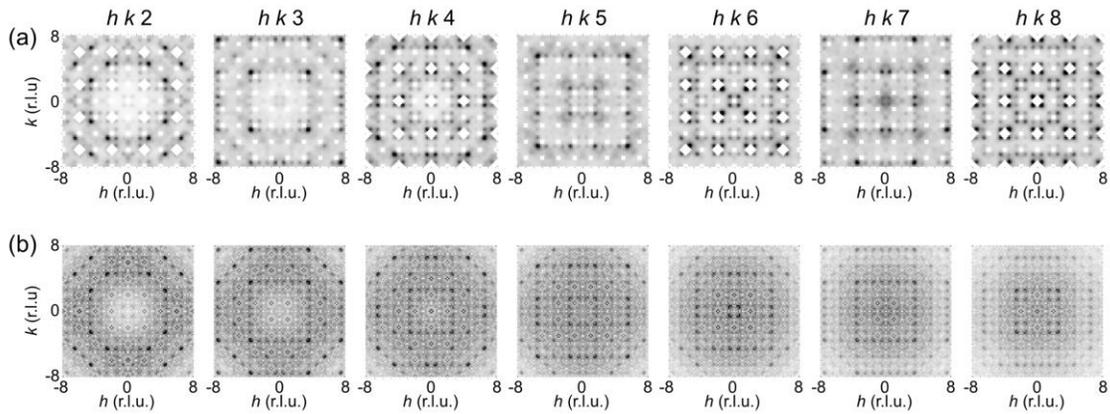

FIG. S9. (a) Experimental XRD data on the $hk$ planes for different $l$ values at 100 K, with Bragg peak intensities removed. (b) Simulated XDS patterns on the $hk$ planes based on the distribution model.

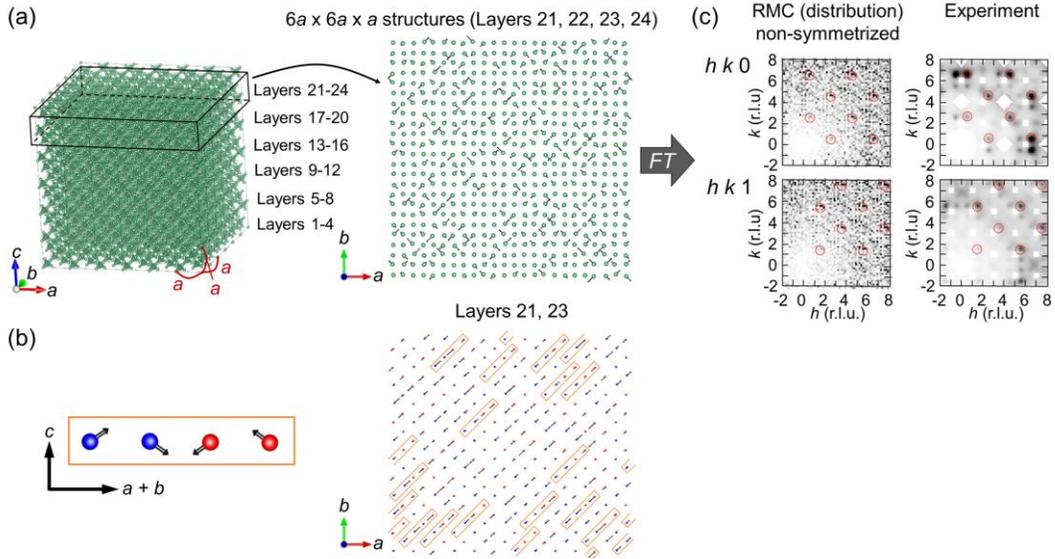

FIG. S10. (a) Results of the RMC simulations based on the distribution model. A four-layer $ab$-plane structure that closely reproduces the experimental XDS pattern is extracted from the $6a \times 6a \times 6a$ supercell. (b) Identical local $Nb_4$ tetramers, enclosed by orange squares, are present in the one-dimensional chains. (c) Non-symmetrized XDS pattern calculated from the RMC simulations, shown alongside the corresponding experimental XDS pattern.

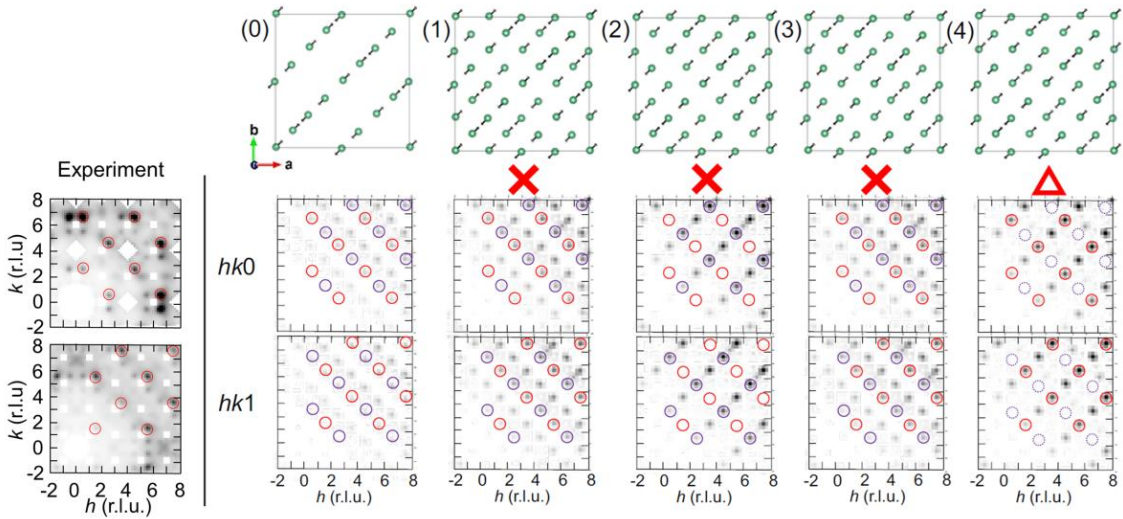

FIG. S11. XDS intensity calculated from several Ising-type Nb displacement models, where only one- or two-layer $ab$-plane structures are considered.

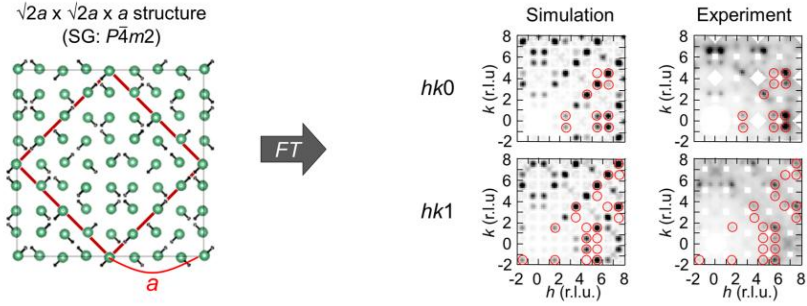

FIG. S12. XDS intensity calculated from the long-range ordered model with the space group $P\bar{4}m2$. All Nb atoms are displaced along the local threefold axes of the cubic lattice with the same magnitude.

**References**

1. N. Metropolis, A. W. Rosenbluth, M. N. Rosenbluth, and A. H. Teller, J. Chem. Phys. **21**, 1087 (1953).
2. T. Proffen and R. B. Neder, DISCUS: a program for diffuse scattering and defect-structure simulation. J. Appl. Cryst. **30**, 171 (1997).